\documentstyle[twocolumn,floats,aps,epsf]{revtex}

\begin{document}

\newcommand{\eg}{\mbox{e.\ g.\ }}
\newcommand{\ie}{\mbox{i.\ e.\ }}

\twocolumn[\hsize\textwidth\columnwidth\hsize\csname@twocolumnfalse%
\endcsname

\title{Orthogonality catastrophe in a one--dimensional system of
correlated electrons}
\author{V.\ Meden$^a$, P.\ Schmitteckert$^b$, and Nic Shannon$^c$}
\address{$^a$Department of Physics, Indiana University, Bloomington, Indiana
47405, U.\ S.\ A.\ \\ 
$^b$IPCMS--GEMME, 23 rue du Loess, 67037 Strasbourg Cedex,
France\\
$^c$Department of Physics, University of Warwick, Coventry CV4
7AL, England}

\date{July 3, 1997}
\maketitle
\draft
\begin{abstract}
We present a detailed numerical study of the orthogonality
catastrophe exponent for a one--dimensional lattice model of spinless
fermions with nearest neighbor interaction using the density matrix
remormalization group algorithm.
Keeping up to $1200$ states per block we achieve a very great accuracy
for the overlap which is needed to extract the orthogonality
exponent reliably.
We discuss the behavior of the exponent for three different kinds
of a localized impurity. 
For comparison we also discuss the non--interacting case.
In the weak impurity limit our results for the overlap 
confirm scaling behavior expected from perturbation theory and
renormalization group calculations. In particular we find that
a weak backward scattering component of the orthogonality 
exponent scales to zero for attractive interaction.
In the strong impurity limit and for repulsive interaction 
we demonstrate that the orthogonality exponent cannot
be extracted from the overlap for systems with up to 100 sites,
due to finite size effects. 
This is in contradiction to an earlier interpretation given by Qin et al.\
based on numerical data for much smaller system sizes. 
Neverthless 
we find indirect evidence that
the backward scattering contribution to the exponent scales to $1/16$
based on predictions of boundary conformal field theory.
\end{abstract}
]

\narrowtext

\smallskip

\section{Introduction}

Anderson's ``orthogonality catastrophe'' is an important and universal
piece of metallic physics with far reaching consequences for the
response functions of many--electron systems.
It is well known that for non--interacting electrons the overlap between
the many--body ground states with and without a local impurity potential
vanishes in the thermodynamic limit
\cite{anderson}.
For a large but finite system of length $L$ the overlap
tends to zero as a power law $L^{-\alpha}$, defining an orthogonality
exponent
(OE) $\alpha$.  This orthogonality catastrophe is closely related to the
x--ray edge problem in metals \cite{nozieres,mahan}, and in this
context the exponent $\alpha$ can often be measured directly.

As was first observed by Anderson \cite{anderson} the OE for a
spherically symmetric impurity in a free electron gas (in arbitrary
dimension) is uniquely determined by the square of the
scattering phase shifts of the
infinite system at the Fermi energy $E_F$.
The proof of this result for more
general potentials is far from trivial \cite{nozieres,yy,haman}.
For spinless one--dimensional (1D) Fermions we can write
\begin{eqnarray}
\label{alphaphase}
\alpha= \frac{1}{2} \left( \frac{\delta_e(E_F)}{\pi} \right)^2 +
\frac{1}{2} \left( \frac{\delta_o(E_F)}{\pi} \right)^2 ,
\end{eqnarray}
with the phase shifts of the even and odd scattering channels.
The remaining problem is the calculation of the phase shifts
for a given impurity potential.
In the following $\delta$ always denotes the phase shift at the Fermi
energy.

In one dimension the phase shifts can generally be expressed in terms
of the reflection and transmission coefficients of an incident wave
packet
\cite{ks2}
\begin{eqnarray}
\label{abexp}
\delta_{e/o}=\frac{1}{2} \left[ \varphi_T \mp \arcsin{ \left\{ |R|
\right\} } \right] ,
\end{eqnarray}
with the phase $\varphi_T$ of the transmission coefficient and the absolute
value $|R|$ of the reflection coefficient taken at the Fermi energy.
If we introduce new phase shifts $\delta_{f/b}$ given by
\begin{eqnarray}
\label{eofb}
\delta_{e/o}=\frac{1}{2} \left( \delta_f \mp \delta_b \right) ,
\end{eqnarray}
we can identify the forward scattering phase shift $\delta_f$ with
$\varphi_T$ and the backward scattering phase shift $\delta_b$ with
$\arcsin{ \left\{ |R| \right\} }$.
Expressed in terms of $\delta_{f/b}$ or $|R|$ and $\varphi_T$,
the OE is given by
\begin{eqnarray}
\label{alphafb}
\alpha & = & \alpha_f + \alpha_b \nonumber \\
& = & \frac{1}{4} \left( \frac{\delta_f}{\pi} \right)^2 +
\frac{1}{4} \left( \frac{\delta_b}{\pi} \right)^2  \nonumber \\
& = & \frac{1}{4 \pi^2}
\left[ \varphi_T^2 + \arcsin^2{ \left\{ |R| \right\} } \right] ,
\end{eqnarray}
defining the forward ($\alpha_f$) and backward ($\alpha_b$)
scattering contributions to the OE.

In the lowest order perturbation theory
in the impurity strength (the Born approximation) it is always
possible to express the scattering phase shifts $\delta_{f/b}$
in terms of the matrix elements of the impurity term $\hat{W}$
in the Hamiltonian.
If we denote the one--particle eigenstates of the
impurity free Hamiltonian with energy $E_F$ by $\left| \pm k_F \right> $,
we obtain
\begin{eqnarray}
\label{fphase}
\left( \delta_f^B \right)^2 & = & \frac{1}{v_F^2} \left|
 \left< k_F \right| \hat{W} \left| k_F \right> \right|^2  \\
\label{bphase}
\left( \delta_b^B \right)^2 & = &  \frac{1}{v_F^2} \left|
\left< -k_F \right| \hat{W} \left| k_F \right> \right|^2  .
\end{eqnarray}

Correlation effects are very important in one dimension, where it is known
that even an infinitesimally small interaction between electrons changes
their low energy properties from those of a Fermi to those of a
Luttinger liquid (LL) \cite{haldane}.
The low energy excitations of a LL are collective
and bosonic, so the concept of phase shifts in fermionic
\mbox{one--(quasi--)particle} wave functions is clearly no longer
applicable, contrary to the situation in interacting Fermi liquids where
this is still possible.
It is therefore interesting to ask whether the orthogonality catastrophe
in a 1D system is also dramatically altered by
electron--electron
interaction.

In this article we analyze the problem by numerically
calculating the overlap
\begin{eqnarray}
\label{overlapdef}
O \equiv \left| \left< \left. E_0 \right| E_0^I \right> \right|
\end{eqnarray}
and the difference in the ground state energies
\begin{eqnarray}
\label{energydifdef}
\Delta E \equiv \left| E_0^I-E_0 \right|
\end{eqnarray}
between the ground states of the system
with ($\left| E_0^I \right>$) and without
($\left| E_0 \right>$)
the impurity for three different kinds of
impurity in a lattice model of spinless interacting Fermions.
We distinguish between the weak and strong impurity limit.
The results are compared with the predictions of perturbation theory
and the application of renormalization group (RG) ideas.
Large system sizes and very accurate calculations are found to be
necessary to reliably determine $\alpha$.  These can be achieved by
using the density matrix
renormalization group algorithm (DMRG) \cite{white,peterdoc}.
This approach
makes it possible to calculate the ground state properties
of an interacting system
with an accuracy comparable to that of exact
diagonalization, but for much larger system sizes.
Using this technique, we are able to work with chains of up to 100
sites, keeping up to 1200 states per block
and performing $7$ finite lattice sweeps in the DMRG procedure.

The behavior of 1D correlated electrons in the presence of a localized
impurity is of general interest because the physics of the problem
manifests itself in a number of different contexts,
\eg the conductivity of quasi 1D ``metallic'' materials \cite{kfc},
tunneling through a constriction in the fractional quantum Hall
regime \cite{moon} and Friedel oscillations in the charge density of
highly anisotropic systems \cite{diplom,egger,peter}.
Given the rapid progress in the fabrication of
quasi 1D metallic systems (highly anisotropic materials or quantum wires)
it should in the future be possible to measure the x--ray response of these
systems directly, and so fit $\alpha$ to experiments.

Considerable progress has been made in the understanding of the
interplay between impurity and interaction in one dimension
since the seminal work of Kane and Fisher \cite{kfc}.
Kane and Fisher discussed the problem of a single impurity
in a 1D wire within the effective low energy field theory of
1D correlated electrons, using a perturbative RG approach.
If one integrates out high momentum degrees of freedom
any {\it weak} $2k_F$ component of the impurity potential increases for
repulsive interaction and scales to zero for attractive ones.
Thus backward scattering of electrons is relevant for repulsive
interaction and irrelevant for attractive interaction.
Kane and Fisher also discussed the
dual problem of a {\it weak} hopping between the two ends
of an open chain.  The strength of this weak link flows to zero for
repulsive interaction, but is relevant for attractive interaction.

It is tempting to conclude that, for repulsive interaction,
the (weak) $2 k_F$ component of the impurity potential scales to
infinity and therefore the open chain fixed point (open chain
interpretation).  However it is not immediately obvious that a single
impurity in a LL really does correspond to a ``cut wire''.

In order to shed more light on this issue Kane and Fisher calculated
the conductivity of a 1D wire with a localized impurity {\it exactly}
for a special value of the LL interaction parameter.
Their result {\it is} consistent with this interpretation of the RG results
\cite{kfc}.
Later authors calculated the conductivity for arbitrary interaction
strength \cite{fendley,lesage}, again obtaining results consistent
with the open chain interpretation.
Further supporting evidence for this picture was given by a numerical
calculation of the conductivity \cite{moon}.
Similar arguments hold in the dual situation of (strong) hopping
between two semi--infinite LL's, attractive interaction, and the
periodic chain fixed point (periodic chain interpretation).

However the
question of how the RG results should be interpreted in the context
of the OE remains controversial
\cite{gogolin,prokofev,kane94,oregprl,oregprb,dispute}.

A number of attempts have been made to calculate
the orthogonality exponent (or equivalently  the x--ray edge exponent)
of a 1D interacting electron gas with strong impurities
in the light of Kane and Fisher's RG results \cite{gogolin,prokofev,kane94}.
All of these approaches are based on the Tomonaga--Luttinger (TL)
Hamiltonian, the effective low energy continuum field theory for most
models of 1D correlated electrons in the absence of impurities.
The TL Hamiltonian is written in terms of distinct left and right moving
fermion fields $\psi_{\pm}(x)$.
The low energy physics of a given microscopic model can be mapped onto
a TL model by determining the LL parameter
$K$ and the renormalized
velocity $v$ of the model in question, in terms of its microscopic
parameters
\cite{johannes}.
Impurity scattering is given by two types of term in the Hamiltonian.
Forward scattering can be written linearly in the density
of right and left moving Fermions.
If only this kind of scattering is present then the calculation of the
overlap within the field theoretical model is straightforward (see
Section \ref{pert}).
Backward scattering couples right and left moving fields
and results in a complicated sine--Gordon like term in the field
theory \cite{kfc}.

To calculate the overlap within the field theory in the presence of
backward scattering further approximations
are needed. To solve the problem for repulsive electron--electron
interaction Gogolin \cite{gogolin} and
Prokof'ev \cite{prokofev} used a strong backward scattering approximation,
replacing the cosine of the boson field by a term quadratic in the field.
This replacement is consistent with the open chain interpretation of the
RG results and leads to a backward scattering contribution
\begin{eqnarray}
\label{onesix}
\alpha_b = \frac{1}{16}
\end{eqnarray}
to the OE, which is completely independent of the strength of
the interaction or the size of the bare backward scattering potential.

This result is frequently interpreted in terms of phase
shifts.  According to the open chain interpretation of Kane and Fisher's
scaling analysis outlined above, the $2 k_F$ component of the impurity is
renormalized
to infinity. This leads to perfect reflection (reflection coefficient
$R=1$) and a backward scattering phase shift of  $\pi/2$.
Introducing this phase shift into Eq.\ (\ref{alphafb})
gives Eq.\ (\ref{onesix}).  As there are no fermionic single--particle
scattering wave functions which are eigenfunctions of the interacting
Hamiltonian, this interpretation only has a meaning if one extends the
concept of phase shifts. We can do this by regarding the phase shift at the
Fermi energy
as the number determining the boundary condition between in-- and outgoing
{\it field operators} according to
$\psi_{\mbox{\scriptsize out}}=\exp{(2i\delta)} \psi_{\mbox{\scriptsize
in}}$.
In this sense the above interpretation {\it does} still have a physical
meaning \cite{affleck94}.

Beside the direct calculation of the overlap,
an alternative method of calculating the OE is given by
boundary conformal field theory (BCFT) \cite{affleck94,zagoskin}.
BCFT relates the properties of the boundary changing operator
associated with the introduction of a single impurity in
a \mbox{(semi--)} infinite 1D interacting Fermi system to the
{\it finite size}
spectrum of an equivalent field theory on a line of length $L$.
For non--interacting electrons and for interacting electrons
with only a forward scattering impurity component
the difference $\Delta E$
of the ground state energies
of the finite size system
with and without the impurity
can be calculated directly from a mode
expansion in the effective low energy field
theory on the line.
In the absence of bound state effects
$\Delta E$ is given by
\begin{eqnarray}
\label{bcftenergy}
\lim_{L \to \infty } L \Delta E=
\lim_{L \to \infty } L \Delta e + 2 \pi v \alpha,
\end{eqnarray}
where $\Delta e$ is a bulk constant and  $v$ the renormalized charge
velocity of the field.
For interacting electrons this result enables us to calculate
$\alpha_f$ directly from numerically determined ground state
energies \cite{affleck94,sebastian}.

An alternative
derivation of Eq.\ (\ref{alphaphase}) for non--interacting electrons
based on Eq.\ (\ref{bcftenergy})
has been given \cite{zagoskin}.
This approach has also been used in conjunction with the Bethe
Ansatz
to obtain
predictions for the orthogonality exponent in lattice systems with
integrable
defects \cite{essler}.
Additionally it is possible to calculate the ground state energy difference
$\Delta E$ in the
presence of an infinitely strong backward scattering potential.
If one {\it assumes} that the scaling of the $2 k_F$ component
of the potential for repulsive electron--electron interaction
finally leads to an open boundary condition fixed
point one again obtains $\alpha_b=1/16$.

It is important to notice that
{\it all} the theoretical results discussed so far are {\it based} on,
or are equivalent to, the open chain interpretation of
the results of the perturbative RG.

Recently the overlap $O$ Eq.\ (\ref{overlapdef}) has been discussed
within the low energy continuum field
theory for the special LL parameter $K=1/2$ corresponding to a strong
repulsive interaction \cite{furusaki,komnik}. At this point it is
possible to calculate
the overlap by bosonization and refermionization.
The result is again consistent with $\alpha_b=1/16$.

In Refs.\ \cite{oregprl} and \cite{oregprb} Oreg and Finkel'stein
discuss the
tunneling density of states near a localized impurity and the x--ray edge
exponent based on the mapping of the TL model onto a Coulomb gas model.
They obtain results in contradiction with results obtained by all other
authors.  Whether their calculations suffer from a flaw in the
anticommutation relations of the Fermion field operators is still
a matter of contention
\cite{dispute,furusaki,komnik,qinI,qinII}.

In Refs.\ \cite{qinI} and \cite{qinII} Qin et al.\ use the DMRG
method to
calculate the overlap for a very special type of localized impurity.
We will later comment on the relation between their results and ours.

The remainder of this article is organized as follows: In Sec.\ \ref{secmod}
we
introduce the model considered, discuss analytical results for
the OE in the non--interacting case, and derive the results for the OE in
lowest
order perturbation theory in the impurity strength, based on the effective
low energy field theory for interacting electrons. In order to obtain the
OE
it is necessary
to extrapolate finite size data for the overlap and the energy
difference to the thermodynamic limit.  This leads to an extrapolation
error.
An additional error is introduced because the DMRG gives only an
approximate
result for the ground state properties. We discuss both errors carefully
in Sec.\ \ref{acc} by calculating the overlap $O$ and $\Delta E$
numerically
for the non--interacting model and comparing with the exact results derived
in Sec.\ \ref{oenoi}.

For non--interacting electrons a straightforward
numerical calculation of $O$ and $\Delta E$ is possible for very large
system sizes, if we take the
Slater determinant structure of the ground states into account.
The comparison of exact results for non--interacting electrons
makes it possible to access the errors involved in the DMRG procedure.
It turns out that this kind of error analysis is crucial
for the interpretation of our results.
In Sec.\ \ref{results}  we present
our numerical results and analyze the data with respect to the expected
scaling behavior. Finally we summarize our results in Sec.\ \ref{sum}.
In the Appendix we briefly describe how to obtain
perturbative results in the  weak
impurity and weak hopping limits.

\section{Models and analytical Results}
\label{secmod}
\subsection{The Hamiltonian}

In our investigation we consider a model of $N$ spinless
Fermions on a 1D lattice of $M$ sites with site dependent
on--site energies $\varepsilon_i$,
nearest neighbor hopping $t$ and interaction $V$, described
by the Hamiltonian
\begin{eqnarray}
\label{eqn1}
H & = & - t \sum_{i=1}^{M}  \left( c_i^{\dag} c_{i+1}^{} + c_{i+1}^{\dag}
c_{i}^{}
\right) + \sum_{i=1}^{M} \varepsilon_i  c_i^{\dag} c_{i}^{}  \nonumber \\
&& + V \sum_{i=1}^{M}  n_i n_{i+1}
+ (b-1) \left[ -t  \left(c_M^{\dag} c_{1}^{} + c_{1}^{\dag}
c_{M}^{} \right) \right] \nonumber \\
&& + (b'-1) \left[ V  n_M n_{1}    \right],
\end{eqnarray}
where $c_i^{(\dag)}$ denotes the Fermion lowering (raising) operator at
the site $i$, and $n_i$ the related density operator. In all sums we
identify $M+1 \equiv 1$. By choosing $b=b'=1$ and
$\varepsilon_i = 0$ for all $i$ we obtain an impurity free
Hamiltonian for interacting Fermions with periodic boundary conditions (PBC).
It is well known that this Hamiltonian displays LL behavior,
and for the case of a half filled band  the LL parameter $K$ and
the renormalized velocity $v$ can be obtained from the finite size
corrections \cite{betheansatz} to the Bethe Ansatz \cite{bethe} solution
\begin{eqnarray}
\label{eqn2}
v & = & t \frac{\pi \sin{(2 \eta)}}{\pi-2\eta} , \\  \label{eqn2a}
K & = & \frac{\pi}{4 \eta}  ,
\end{eqnarray}
where $\eta$ parameterizes the interaction
\begin{eqnarray}
\label{eqn3}
V=-2t \cos{(2\eta)} .
\end{eqnarray}
For $V>0$ (repulsive interaction), $K<1$, and for $V<0$ (attractive
interaction) $K>1$.
In this article we will focus on the half filled case.
According to the LL theory of 1D correlated electrons \cite{haldane} $K$
and
$v$ completely determine the low--energy behavior of the Hamiltonian.
It is known that in the absence of impurities the model described by
Eq.\ (\ref{eqn1}) undergoes a transition into a charge
density wave ground state at $V/|t|=2$, and into a phase separated state
at $V/|t|=-2$. We are therefore limited to parameters $V/|t|$ within
$(-2,2)$ leading to $K \in (1/2,\infty)$.

By varying the parameters $\varepsilon_i$, $b$ and $b'$ we can introduce
different kinds of localized impurity to the Hamiltonian Eq.\
(\ref{eqn1}).
In this article we consider three kinds of impurity.  Setting
$\varepsilon_1=\varepsilon$, $\varepsilon_i=0$ for $i\neq 1$ and
$b=b'=1$ creates a {\it site impurity} on one of the lattice sites.
Without loss
of generality we have chosen the site $1$ as the impurity site.
Physically one might think of this kind of term as an impurity ion
sitting on one of the sites of the regular lattice, or as the
core hole generated by removing one (or more) of the core electrons of the
atom at site $1$ (x--ray edge problem). In this article we will focus
on repulsive impurities ($\varepsilon>0$) to avoid complications due to
bound
states.
A {\it hopping
impurity} is obtained by choosing $\varepsilon_i=0$ for all $i$, $b'=1$ and
$0 \leq b <1$ and might be created by an interstitial between lattice
sites
$M$ and $1$ which changes the hopping amplitude but not the
interaction between them.

Qin et al.\ \cite{qinI,qinII}  recently discussed a third type of
impurity. These authors worked with the Heisenberg XXZ spin 1/2 chain,
which can be mapped onto a model of spinless Fermions by a Wigner--Jordan
transformation \cite{lieb}. In terms of the Heisenberg model, their impurity
is generated by weakening the coupling of the $x$--, $y$-- and
$z$--components
of the spin between sites $M$ and $1$ by an equal amount. In the
Fermion basis, this impurity corresponds to an exotic combination
of modified hopping and interaction between the sites $M$ and $1$,
accompanied by site impurities at the sites $M$ and $1$.
Their model is equivalent to  Eq.\ (\ref{eqn1}) with
$\varepsilon_1=\varepsilon_M = -(b-1) V /2 $,
$\varepsilon_i=0$ for all the other $i$, $b=b'$ and $0 \leq b <1$.
For comparison we will also discuss this type of impurity, and
denote it a {\it bond impurity.}
It is equivalent to the hopping impurity in the non--interacting
limit $V=0$.

The ground state of the Hamiltonian Eq.\ (\ref{eqn1}) without
impurities is {\it degenerate} for even particle number $N$.
There are several ways of overcoming the problems associated with
this degeneracy and (physical) even--odd effects.
One straightforward solution is to calculate $O$ and $\Delta E$ only for
odd $N$.  However, in order to have access to more data points we have
chosen a different approach in the majority of our numerical calculations.
For even $N$ we altered the hopping between the sites
$M$ and $1$ from $t$ to $-t$.  This lifts the degeneracy of
the ground state, and prevents even--odd effects irrelevant
to the orthogonality catastrophe considered in this article.
One can implement this in the Hamiltonian by the direct substitution
\begin{eqnarray}
\label{repl}
(b-1)  \longrightarrow \left( - \exp{ \left\{ i \pi \sum_{j=1}^{M} c_j^{\dag}
c_j \right\} } b-1 \right)
\end{eqnarray}
in the second line of Eq.\ (\ref{eqn1}).  If one works with the Heisenberg
Hamiltonian and imposes PBC's for the spins this additional phase factor
emerges as a natural consequence of the Wigner--Jordan transformation
\cite{lieb}.

\subsection{The orthogonality exponent for non--interacting electrons}
\label{oenoi}
As shown by Eq.\ (\ref{alphafb}), we can determine the OE $\alpha$
for non--interacting electrons by solving the one--particle
scattering problem and calculating $|R|$ and $\varphi_T$.
For the types of impurities considered in this article this is
straightforward and can be done following e.\ g.\ Ref.\
\cite{ks1}. For the site impurity we obtain
\begin{eqnarray}
\label{sialpha}
\alpha= \frac{1}{2 \pi^2} \arcsin^2{ \left\{
\frac{\varepsilon}{\sqrt{4 t^2+\varepsilon^2 - E_F^2}} \right\} } .
\end{eqnarray}
The forward and backward scattering contributions $\alpha_{f/b}$
are both given by one--half of $\alpha$, as $\varphi_T= - \arcsin{|R|}$,
and $|R|$ is given by the argument of the arcsine in Eq.\
(\ref{sialpha}).
In preparation for the discussion of the OE in terms of an effective
low energy field theory in the next section, we also calculate
$\delta_{f/b}$ in the Born approximation and at half filling.
This will be necessary in order to
relate the impurity parameters of the microscopic lattice model to the
impurity potentials in the field theory.
We denote the value of the OE obtained from the
Born approximation as $\alpha^{B}$.
Following Eqs.\ (\ref{fphase}) and (\ref{bphase}) we obtain
$( \delta_f^B)^2= ( \delta_b^B )^2
=(\varepsilon/v_F)^2$ with the non--interacting Fermi velocity
$v_F = 2|t|$. Inserting this result into Eq.\ (\ref{alphafb}) leads
to the same result as an expansion of Eq.\ (\ref{sialpha}) to lowest
order in $\varepsilon$
\begin{eqnarray}
\label{baalphai}
\alpha^B = \frac{1}{2\pi^2} \left(
\frac{\varepsilon}{v_F} \right)^2 .
\end{eqnarray}
The next order corrections to $\alpha^B$ are of the order
$(\varepsilon/v_F)^4$.

To simplify the calculation in the case of the hopping impurity we
concentrate on the half filled band with $E_F=0$.  In this case
\begin{eqnarray}
\label{bialpha}
\alpha= \frac{1}{4 \pi^2} \arcsin^2{ \left\{
\frac{ \beta - \beta^2/2 }{1-\beta +\beta^2/2}
\right\} } ,
\end{eqnarray}
with $\beta = (1-b)$.
For the hopping impurity at half filling
the {\it forward scattering contribution vanishes,} e.\ g.\ we obtain
\begin{eqnarray}
\label{forzero}
\delta_f \equiv \varphi_T = 0
\end{eqnarray}
and $|R|$ is again given by the argument of the
arcsine in Eq.\ (\ref{bialpha}).
For a weak impurity (small $\beta$) an expansion leads to
\begin{eqnarray}
\label{weak}
\alpha^B=\beta^2/(4 \pi^2)
\end{eqnarray}
and $\left(\delta_b^B \right)^2 = \beta^2$.
In contrast to the site impurity the next order corrections to the Born
approximation are of the order $\beta^3$ and therefore more
important than for the site impurity. This observation will be important
in what follows.
As the impurity strength in the
lattice model is not determined by a local impurity potential,
the calculation of the backward scattering
phase shift in the Born approximation following Eq.\ (\ref{bphase}) is
a little bit more involved, but also leads to Eq.\ (\ref{weak}).

\subsection{Bosonization and perturbation theory in the impurity strength}
\label{pert}
In the low energy, long wavelength limit, the impurity free Hamiltonian
can be written in terms of a bosonized continuum field
theory \cite{johannes}.
The new effective Hamiltonian is given by
\begin{eqnarray}
\label{fieldh}
{\mathcal H}_0 = \frac{v}{2 \pi} \int_{-L/2}^{L/2}
dx \left\{K \pi^2
\Pi^2(x) + \frac{1}{K}
\left( \frac{\partial \Phi(x)}{\partial x} \right)^2
\right\} ,
\end{eqnarray}
and the canonical conjugate bosonic fields $\Phi(x)$ and $\Pi(x)$ can be
expressed in terms of the densities $\rho_{\pm}$ of right
and left moving Fermions \cite{johannes}
\begin{eqnarray}
\label{thefields1}
\Pi(x) & = & \rho_+(x) - \rho_-(x)  , \\
\label{thefields2}
\Phi(x) & = & -i \frac{\pi}{L} \sum_{q \neq 0} \frac{1}{q} e^{-iqx}
\left[ \rho_+(q) + \rho_-(q) \right]
\nonumber \\
&& - \frac{\pi x}{L}
\left[ \hat{N}_+ + \hat{N}_- \right],
\end{eqnarray}
where the number operators $\hat{N}_{\pm}$ measure the particle
number with respect to a fixed ground state.
In the non--interacting model $K=1$ and  the velocity $v$ is given
by the non--interacting Fermi velocity $v_F$.

A forward scattering impurity can be expressed  as
\begin{eqnarray}
\label{hfore}
{\mathcal W}_f = \int_{-L/2}^{L/2} W_f(x) \rho(x) ,
\end{eqnarray}
with the total density $ \rho(x) = \left[ \rho_+(x) + \rho_-(x) \right] $
and the local potential $W_f(x)$.  Eq.\ (\ref{thefields2}) implies
\begin{eqnarray}
\label{rhoder}
\rho(x)= - \frac{1}{\pi} \frac{\partial \Phi(x)}{\partial x} .
\end{eqnarray}
Since ${\mathcal W}_f$ is linear in the $\rho_{\pm}(q)$, the ground states
of ${\mathcal H}_0$ and ${\mathcal H}_0 + {\mathcal W}_f $ are related
by a unitary transformation and the overlap can be calculated
explicitly
\begin{eqnarray}
\label{forewardo}
O_f = \exp{ \left\{- \frac{1}{2 \pi L} \frac{K}{v^2} \sum_{q>0}
\frac{|\tilde{W}_f(q)|^2}{q}      \right\}  } .
\end{eqnarray}
Taking the limit $L \to \infty$ we can determine the OE
\cite{diplom,forrefs}
\begin{eqnarray}
\label{expfore}
\alpha_f= \frac{K}{4 \pi^2} \frac{|\tilde{W}_f(0)|^2}{v^2}.
\end{eqnarray}
Anderson's orthogonality catastrophe is therefore still present in
an interacting system with forward scattering, but the OE is modified.
Eq.\ (\ref{expfore}) trivially reproduces the non--interacting limit,
and we conclude from comparison
with Eq.\ (\ref{fphase}) that the field theoretical
description always gives the Born approximation
in $W_f$ for the OE \cite{schotte}.
We therefore anticipate that for $V \neq 0$, Eq.\ (\ref{expfore})
also holds only for weak forward scattering.

This raises the general question of how the specific impurities
(\eg bond, site, hopping) of a given microscopic model should be
mapped onto the impurity potential $\tilde{W}_f(q)$ of the effective
field theory.  Comparing Eq.\ (\ref{expfore}) for $V=0$ with
Eq.\ (\ref{alphafb}) we find
\begin{eqnarray}
\label{identi}
\frac{|\tilde{W}_f(0)|^2}{v_F^2} = \delta_f^2 ,
\end{eqnarray}
where $\delta_f$ is the exact non--interacting
forward scattering phase shift of the
underlying microscopic model.  Within this mapping
we obtain from the field theoretical Born approximation the
exact result for $\alpha_f$ of the microscopic model.
But it is {\it not} clear whether this identification
also gives a systematic improvement in the interacting case.
From a systematic point of view we can only
identify the impurity parameters in lowest order
in perturbation theory
$[\tilde{W}_f(0)/v_F]^2=(\delta_f^B)^2$,
where  $\delta_f^B$ is the non--interacting microscopic
forward scattering phase shift in the Born approximation.

For small $|\tilde{W}_f(0)|$ we can therefore  write
\begin{eqnarray}
\label{wf}
\alpha_f^B(V)
=K \frac{v_F^2}{v^2} \alpha_f^B (V=0).
\end{eqnarray}
Having calculated $ \alpha_f^B (V=0)$
for our microscopic site impurity in Eq.\ (\ref{baalphai}),
$\alpha_f^B(V)$ is determined by Eq.\
(\ref{wf}).
We will later compare the values of $\alpha_f$ predicted by this equation
with our numerical results for weak site impurities.

To the best of our knowledge
there are no analytical results for the case of strong forward scattering
and \mbox{$V\neq 0$}.

It is important to note that the inclusion of the forward scattering term
Eq.\ (\ref{hfore}) in the Hamiltonian breaks its particle--hole symmetry.
As the lattice Hamiltonians
Eq.\ (\ref{eqn1}) for the {\it hopping} and {\it bond}
impurities at half filling are particle--hole symmetric, in this 
special case 
the mapping of the microscopic model onto an effective field
theory {\it cannot} give rise to such a term.  In this sense
the forward scattering contribution to the OE vanishes for
interacting electrons as well, which simplifies the interpretation
of some of our results.

For purposes of comparison with numerical data for finite size systems,
it is interesting to evaluate Eq.\ (\ref{forewardo}) with finite $L$.
Provided that we can expand $|\tilde{W}_f(q)|^2$ in a power series in $q$,
the Euler summation formula \cite{abramowitz} gives
\begin{eqnarray}
\label{finitecorr}
O=\exp{ \left\{- \alpha_f \ln{(L)} + a_0 + a_1 L^{-1} + a_2 L^{-2} + \ldots
 \right\}  } ,
\end{eqnarray}
with impurity strength and interaction dependent constants $a_i$.
In the Appendix we will argue that in the presence of backward scattering
the finite size corrections for $V=0$ are also given by integer powers of
$1/L$.

The backward scattering contribution to the low energy field theory is
given by
\begin{eqnarray}
\label{hback}
{\mathcal W}_b = \int_{-L/2}^{L/2} W_b(x) \left[ \psi_+^{\dag}(x)
\psi_-^{}(x) + \mbox{h.c.} \right] ,
\end{eqnarray}
where the field operators $\psi_{\pm}(x)$ can be expressed in terms
of the Boson fields $\Pi(x)$ and $\Phi(x)$ \cite{haldane,johannes}.
In the Appendix we derive an
expression for the overlap perturbatively in the backward scattering
potential.
If only backward scattering is present \cite{diplom}
\begin{eqnarray}
\label{backwardo}
O_b=1-a_1(V) \frac{W_b^2}{v_F^2} - a_2(V) \frac{W_b^2}{v_F^2}
L^{2-2K} + \ldots ,
\end{eqnarray}
where the $a_i(V)$ are interaction dependent constants, $W_b$ is a measure
of the backward scattering strength and the dots denote
terms of second order in $W_b/v_F$ falling off as $1/L$ or faster
and terms of higher order in $W_b/v_F$.
This result is consistent with the RG
equations given in Ref.\ \cite{kfc}, and is related to the anomalous time
dependence of the core--hole Green's function discussed in
Ref.\ \cite{gogolin}.
It is not obvious from Eq.\ (\ref{backwardo}) that
we will recover the non--interacting form
\begin{eqnarray}
\label{backwardonointer}
O_b=1-a_1  \frac{W_b^2}{v_F^2}  - \alpha_b^{\scriptsize B}
\ln{(L)} + \ldots
\end{eqnarray}
in the limit $K \to 1$.
By taking the explicit $V$ dependence of the coefficients $a_i(V)$ into
account, we show in the Appendix that this is in fact the case.

For repulsive interaction ($K<1$), the expansion Eq.\ (\ref{backwardo})
breaks down at
large system sizes, even if the bare backward scattering
potential is weak.
The same is true in the non--interacting case, but the logarithmic
divergence there is replaced by a faster power law divergence. 
As we do not find the usual
behavior Eq.\ (\ref{backwardonointer}) in the interacting case, we
expect that the orthogonality catastrophe in the presence of backward
scattering and interaction is drastically altered.
For attractive interaction ($K>1$) and in the lowest order perturbation
theory $O_b$ tends to a {\it constant}.
Provided that this holds also in higher orders, there is
{\it no orthogonality catastrophe},
\ie the two ground states are not orthogonal to each other.
Where the bare $|W_b|$ is small, we expect to find the scaling behavior
Eq.\ (\ref{backwardo}) in
our numerical data for the overlap, for intermediate
system sizes.  Terms of higher order in $1/L$
will be important for very small systems.

As in the forward scattering case a systematic mapping between the
microscopic backward scattering parameter and $W_b$ is only possible
within perturbation theory. In lowest order we can identify
$[W_b(0)/v_F]^2=(\delta_b^B)^2$, where $\delta_b^B$ is the non--interacting
microscopic backward scattering phase shift in the Born approximation.

In Ref.\ \cite{kfc} Kane and Fisher discuss
the strong impurity limit by introducing a weak hopping in an open chain
and mapping this problem to a field theory dual to Eq.\ (\ref{fieldh}).
This mapping makes it possible to treat the problem perturbatively in
the weak hopping. In our case this corresponds to a calculation of the overlap
between the state with the weak hopping (the large hopping
or bond impurity) and the chain with open boundary conditions (OBC).
To distinguish this overlap from
the one discussed previously, we denote it $O^o$, and the related OE
$\alpha^o$.
As discussed in the Appendix we can calculate $O_b^o$ perturbatively in
the weak hopping limit in a similar way as $O_b$ and the result
is given by an expression equivalent to
Eq.\ (\ref{backwardo}) but with $K$ replaced by $1/K$
\begin{eqnarray}
\label{backwardodual}
O_b^o=1-\tilde{a}_1(V)  \frac{t_b^2}{t^2}- \tilde{a}_2(V)
 \frac{t_b^2}{t^2} L^{2-2/K} + \ldots ,
\end{eqnarray}
where $t_b$ is now a measure of the weak hopping.
The roles of repulsive and attractive interaction are interchanged
and $O_b^o$ tends to a constant for repulsive interaction.
We will verify Eq.\ (\ref{backwardodual}) numerically
in Sec.\ \ref{weakhopsec}.

BCFT predicts that $\alpha_b$ between the
OBC and the PBC ground states is $1/16$, and that this result is
independent of $V$ \cite{affleck94}.
Bearing in mind all the results discussed so far, it is therefore tempting
to conclude that $\alpha_b(V<0)=0$ and  $\alpha_b(V>0)=1/16$ independent
of the bare impurity strength.  We want
to emphasize that these generalizations of the above perturbative results
are consistent with the periodic and open chain interpretation of
the perturbative RG results, but
need to be verified.  For our models we will discuss this question
in Sec.\ \ref{results}.

In the next section we will discuss the two major sources for errors in
our numerical calculation of the OE: The extrapolation error, and the
error due to the approximative nature of the DMRG. In this section we
will also introduce the method used to analyze the numerical data for
$O$ and $\Delta E$ in order to obtain $\alpha$.

\section{Accuracy of the numerical results}
\label{acc}

\subsection{Extrapolation to infinite system size}
\label{secextra}

For {\it non--interacting} electrons, the ground states
with and without the impurity are Slater determinants of single
electron states, and the overlap can be written as the determinant
of a matrix whose elements are given
by the overlaps between these single electron wave functions
\cite{anderson}.  The single particle Hamiltonian can be diagonalized
numerically, and it is therefore possible to calculate $O$
for quite large system sizes with an error limited only by machine
accuracy.
To discuss the finite size behavior of
the overlap and hence the OE, we calculated $O$ and $\Delta E$ for
non--interacting systems with $M$ up to $1200$.

We expect $O$ to be given by Eq.\ (\ref{finitecorr}) with
$\alpha_f$ replaced by $\alpha$.
We can determine
$\alpha$ by calculating $O$ for different system sizes $M[i]$, 
$i=1,2,3, \ldots$,
taking the centered differences
\begin{eqnarray}
\label{centerddef}
\alpha^O[i]= - \frac{ \ln{ \left\{ O(M[i+1]) \right\} }  -
\ln{ \left\{ O(M[i-1]) \right\} }}{  \ln{ \left\{ M[i+1] \right\} }
-     \ln{ \left\{ M[i-1] \right\} } }
\end{eqnarray}
and extrapolating by fitting a power law $a+bx+cx^2$ with $x=1/M$ in some
window of small $1/M$.

Alternatively we can calculate $\alpha$ using the results of BCFT
Eq.\ (\ref{bcftenergy}). For non--interacting electrons BCFT
predicts that the finite size corrections are given by integer powers
of $1/M$ \cite{affleck94,zagoskin}. Thus we calculate $\Delta E$
for different $M[i]$,
take the centered differences
\begin{eqnarray}
\label{centerddefa}
\alpha^{\Delta E}[i]= \frac{1}{2 \pi v} \left|  \frac{  \Delta E(M[i+1])
-
 \Delta E (M[i-1]) }{  \left( 1/M[i+1] \right)
-     \left(1/ M[i-1] \right) } \right|
\end{eqnarray}
and once again fit the data with a quadratic polynomial in $1/M$.

\begin{figure}[tb]
\begin{center}
\leavevmode
\epsfxsize \columnwidth
\epsffile{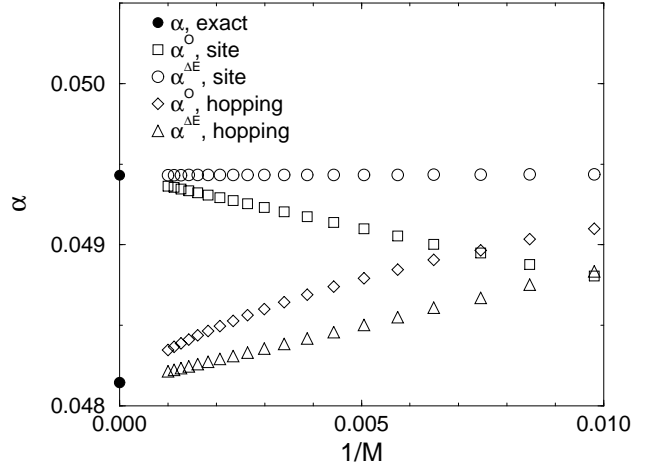}
\caption{The non--interacting orthogonality exponent extrapolated from
a finite size scaling of the overlap and the energy difference for a site
and a hopping impurity with $\varepsilon/|t|=3$ and $b=0.1$.}
\label{fig1}
\end{center}
\end{figure}

Fig.\ \ref{fig1} shows $\alpha^O[i]$ and $ \alpha^{\Delta E}[i]$ for
the site and hopping impurities in the strong impurity limit with
$\varepsilon/|t| = 3$ and $b=0.1$ at half filling.
The results discussed in this section
are quite general and not limited to the case of strong
impurities.  From Fig.\ \ref{fig1} it is obvious that we can confirm
the prediction of BCFT Eq.\ (\ref{bcftenergy}).
The exact values of $\alpha$ obtained
from Eqs.\ (\ref{sialpha}) and (\ref{bialpha}), and the absolute
extrapolation
errors of a finite size scaling in different $1/M$ windows are given in
Table
\ref{table1}.
Fitting only the data with $M[i] \in [200,1200]$ we obtain
numerical values for $\alpha$ with a very high accuracy.
For $M[i] \in [50,200]$
which is comparable to the range accessible by the DMRG the relative
extrapolation error is between $10^{-2}$ and $10^{-5}$, depending on the
impurity type and whether the extrapolation is of $\alpha^O[i]$ or
$ \alpha^{\Delta E}[i]$.  Generally the accuracy is at least one
order of magnitude better for the site impurity than it is for the
hopping impurity. We therefore conclude that while the hopping impurity has
the advantage of generating only backward scattering, its finite size
corrections are much more important than those for the
site impurity.  As there is no reason to believe that
the extrapolation error for interacting electrons is smaller than that
for non--interacting electrons this observation will be important in
what follows.  We also notice that the value of $\alpha$ obtained by the
extrapolation
of $ \alpha^{\Delta E}[i]$ is at least one order of magnitude better than
the one from $\alpha^O[i]$.

\subsection{Accuracy of the DMRG}
\label{secaccdmrg}

The DMRG algorithm is a real space blocking method that works particularly
well for the calculation of the ground state energies and wave functions
of 1D electronic lattice models.
The accuracy of the method is principally limited
by the number $m$ of states that
are kept after the diagonalization of two joint blocks.
A compromise must be reached between keeping $m$ small in order
to reduce the numerical effort, and retaining enough states to
give a good description of the system built out of the blocks.

\begin{figure}[tb]
\begin{center}
\leavevmode
\epsfxsize \columnwidth
\epsffile{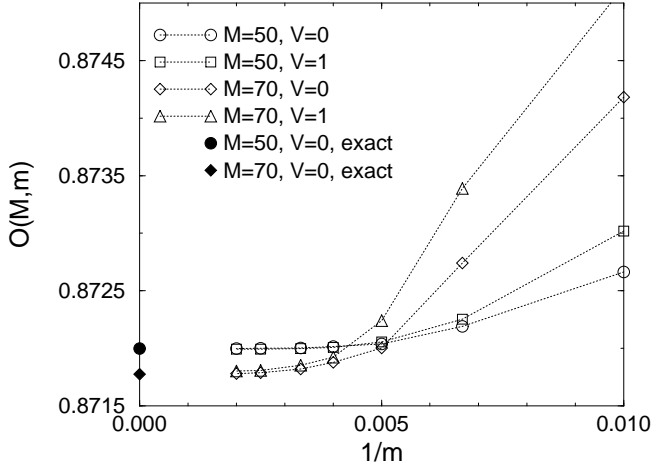}
\caption{The overlap as a function of $1/m$. The full symbols are the
exact result for $V=0$ obtained by diagonalizing the one--particle
Hamiltonian. For comparison, the full diamond is shifted by $0.014$.
The open symbols are the DMRG results. The circles are the
overlap for $M=50$, $V=0$ and a hopping impurity with $b=0.1$. The
squares are the data for $M=50$, $V/|t|=1$ and a bond impurity with
$b=0.1$.
The data are shifted by $0.03026$. The diamonds and triangles are the
data for the same parameters but $M=70$. The diamonds are shifted by
$0.014$
and the triangles by $0.0474$. In the legend $V$ is measured
in units of $|t|$.}
\label{fig2}
\end{center}
\end{figure}

As the non--interacting case is not
a special situation for DMRG,
a comparison with the ``exact''
results for $O$ calculated as in Sec.\ \ref{secextra} gives us a reliable
measure of its error. We therefore used the DMRG to numerically 
calculate $O(M,m)$ for the site and 
hopping impurity with different impurity parameters, different $M$ and
$m$. For the same impurity parameter and $M$ we calculated
$O(M,\infty)$ as in Sec.\ \ref{secextra}. 
As a generic example we show in Fig.\ \ref{fig2}
$O(M,m)$ for a hopping impurity with $V=0$ and $b=0.1$
as a function of $1/m$ for $M=50$ and $M=70$. 
Additionally the figure
shows $O(M,m)$ for the bond impurity with $V/|t|=1$ and $b=0.1$. These are
the parameters used in Ref.\ \cite{qinI}.
To present the curves for the different $M$ and $V$ in one figure
the data for $V=0$, $M=70$ and $V/|t|=1$, $M=50,70$ are shifted
as indicated in the caption.
It is very important to notice that for all $M$ the overlap approaches
the $m=\infty$ limit from {\it above} and that $O(M,m)- O(M,\infty)$
is increasing very fast with $M$.
To calculate $\alpha$ from $O(M,m)$
we have to take a numerical derivative (see Eq.\ (\ref{centerddef})).
If we fix $m$ and increase $M$, $O(M,m)$ is too
large and $O(M,m)- O(M,\infty)$ increases exponentially with $M$.
The numerical differentiation in Eq.\ (\ref{centerddef})
reinforces this effect and we recover values of
$\alpha[i]$ which are systematically too small. For large $M[i]$ this
leads to a significant error in $\alpha[i]$.

Although we do not know
$O(M,\infty)$ for $V \neq 0$ it is obvious from Fig.\ \ref{fig2}
that the behavior
of $O(M,m)$ is quite similar to that for $V=0$.

The same problem occurs for the other types of impurity and
in calculations of $\alpha$ from $\Delta E$, as $\Delta E(M,m)$
always approaches $\Delta E(M, \infty)$ from below. The absolute
error in $\Delta E$ is of the same order of magnitude as the one in $O$.

The best way to overcome this difficulty would be an extrapolation
from finite $m$ to
$m=\infty$. Unfortunately the functional dependence of $O(M,m)$ and
$\Delta E(M,m)$ on $m$ is not known. The numerical data for $O(M,m)$
and $\Delta E(M,m)$ show signs of a power law dependence on $1/m$ with
an exponent of the order of $4.5$, 
but we found it impossible to extract a convincing extrapolation law.
By choosing very large $m$ we instead ensure
that the systematic error is always insignificantly small.
To this end we keep  $m=600$ states per block for
$M \in [6,48]$, $m=800$ for $M \in [50,58]$, $m=1000$ for $M \in [62,80]$,
$m=1100$ for $M \in [84,90]$, and $ m=1200$ for $M=100$ for all data sets
shown in the next section. For these $m$  the
value of $O(M,m)- O(M,\infty)$ for $M=100$ is smaller than $10^{-6}$,
and it is even less for smaller values of $M$.  As we will see in
Sec.\ \ref{results}, this level of accuracy is needed to obtain
meaningful results.

As can be seen in Fig.\ \ref{fig2} the error for the interacting
data is larger than for the non--interacting ones. This observation
holds for all $M$, $\Delta E$ and the other types of impurities.

\section{Numerical results for interacting electrons}
\label{results}

In this section we present and interpret numerical results
for the three chosen types of local impurities. In the discussion we
will distinguish between the weak impurity limit where we expect
to find results  given by the Eqs.\ (\ref{wf}) and (\ref{backwardo}), the
weak hopping limit where we calculate the overlap with the
OBC ground state and expect Eq.\ (\ref{backwardodual}) to hold, and
the strong impurity limit.

\subsection{Weak impurities}
\label{weaksec}

\begin{figure}[tb]
\begin{center}
\leavevmode
\epsfxsize \columnwidth
\epsffile{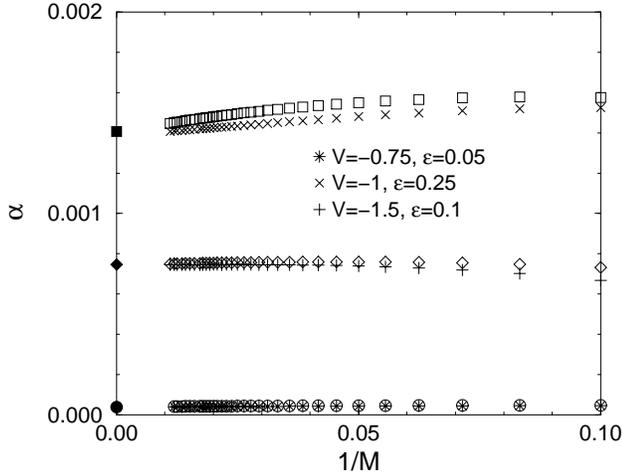}
\caption{The orthogonality exponent for weak site impurities and attractive
interaction. The stars
are $\alpha^O$ and the circles are $\alpha^{\Delta E}$
for $V/|t|=-0.75$, $\varepsilon/|t|=0.05$, the crosses are
$\alpha^O$ and the squares are $\alpha^{\Delta E}$ for
$V/|t|=-1$, $\varepsilon/|t|=0.25$ and the plus signs are
$\alpha^O$ and the diamonds are $\alpha^{\Delta E}$
for $V/|t|=-1.5$, $\varepsilon/|t|=0.1$.
The full symbols represent the related $\alpha_f^B$.
In the legend  $V$ and $\varepsilon$ are measured in units of $|t|$.}
\label{fig3}
\end{center}
\end{figure}

For weak impurities and attractive interaction perturbation theory
predicts that the backward scattering
contribution to the overlap tends to a constant
(see Eq.\ (\ref{backwardo})).  The remaining $M$ dependence of $O$ is given
by the forward scattering contribution. To check this, and Eq.\
(\ref{wf}) for
$\alpha_f^B(V)$, we calculated $O$ and
$\Delta E$ for several
values of $V<0$ and weak site impurities.  From these data we found
$\alpha^{O}[i]$ and $\alpha^{\Delta E}[i]$  as determined by
Eqs.\ (\ref{centerddef}) and (\ref{centerddefa}).  Fig.\ \ref{fig3}
shows the results for the parameter sets $V/|t|=-0.75$,
$\varepsilon/|t|=0.05$,
$V/|t|=-1$, $\varepsilon/|t|=0.25$ and $V/|t|=-1.5$,
$\varepsilon/|t|=0.1$.
{\it Assuming} that at large $M$ the finite
size scaling is given by integer powers of $1/M$ Eq.\ (\ref{finitecorr})
and not an anomalous scaling of the form
Eq.\ (\ref{backwardo}), we can extrapolate
the data by fitting polynomials up to second order in $1/M$.
As the fitting
works very well this procedure is justified.
In Table \ref{table2}
we list $\alpha^{\mbox{\scriptsize exact}}(0)$,
$\alpha^B(0)$,
$\alpha_f^B(V)$,
and the extrapolated $\alpha(V)$ for a fit with $1/M \in [0.01,0.03]$.
Taking into account the extrapolation error and the fact that the
Born approximation is
even in the non--interacting case only an approximation (see Table
\ref{table2}), we conclude
that our data are consistent with the predictions of perturbation
theory and the RG.
There is no backward
scattering contribution to $\alpha$ and $\alpha_f(V)$ is given
by $\alpha_f^B(V)$.  We do not believe that the
difference
between the analytic values of $\alpha_f^B(V)$
and the extrapolated $\alpha$ can be explained by a residual backward
scattering contribution. The difference is much smaller than the backward
scattering contribution for $V=0$ [as discussed following Eq.\
(\ref{sialpha})
$\alpha_b (V=0)=\alpha (V=0)/2$] and for some of the parameter sets
the extrapolated $\alpha$ is smaller than the analytical
$\alpha_f^B(V)$.
As $\alpha^{O}[i]$ and $\alpha^{\Delta E}[i]$ tend
to the same limit as $M \to \infty$ the data are consistent with
the results of BCFT (Eq.\ (\ref{bcftenergy})).

We would like to emphasize that for weak impurities it is important
to calculate $O$ and $\Delta E$ with a very high numerical accuracy.
The difference between $O$ and $1$ is very small, \eg for
$V/|t|=-0.75$ and $\varepsilon/|t|=0.05$ this difference is of the
order of $10^{-4}$. Therefore results are only meaningful if
$O$ and $\Delta E$ have a much smaller error.
As discussed in the last section the error in $O$
and $\Delta E$ in our calculations is less than $10^{-6}$.

\begin{figure}[tb]
\begin{center}
\leavevmode
\epsfxsize \columnwidth
\epsffile{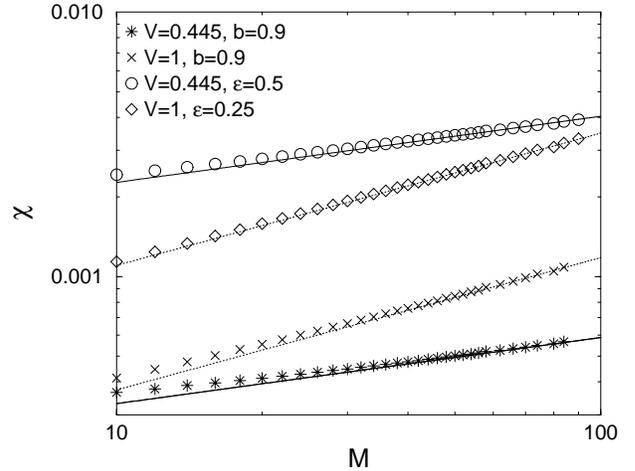}
\caption{Scaling behavior of the overlap for weak impurities and repulsive
interaction. The symbols are the DMRG results for parameters as indicated in
the legend.
The solid lines are power laws with
exponent $2-2K=0.25$  and the dotted lines have the exponent $2-2K=0.5$.
In the legend $V$
and $\varepsilon$ are measured in units of $|t|$.}
\label{fig4}
\end{center}
\end{figure}

The next goal is to confirm Eq.\ (\ref{backwardo}) for weak hopping
impurities, \ie  small $1-b$, with attractive and repulsive
interaction and
weak site impurities with repulsive interaction.  In the case of the hopping
impurity there is only backward scattering, and $O$ is given by $O_b$. For
site impurities we have to distinguish between the conventional weak
forward
scattering contribution to $O \sim - \alpha_f(V) \ln{(M)}$
and the anomalous scaling
given by Eq.\ (\ref{backwardo}). We have already confirmed that the
analytical
result Eq.\ (\ref{wf}) for $\alpha_f^B(V)$
is a good approximation for small $\varepsilon/|t|$.  Therefore we can
extract the anomalous behavior by subtracting $- \alpha_f^B(V) \ln{(M)}$.
In the following $O$ always stands for the anomalous contribution.

As the constant $a_1(V)$ in Eq.\ (\ref{backwardo}) is unknown we
analyze $O$ by calculating centered differences of $O$ and expect
from  Eq.\ (\ref{backwardo})
\begin{eqnarray}
\label{plottingway1}
\chi[i] & \equiv & -
\frac{O(M[i+1])-O(M[i-1])}{  \ln{ \left\{ M[i+1] \right\} }
-     \ln{ \left\{ M[i-1] \right\} } } \nonumber \\
& = & a_2(V)  \frac{W_b^2}{v_F^2} (2-2K) M^{2-2K} + \ldots \,\, .
\end{eqnarray}
In lowest order in perturbation theory $W_b/v_F$ for the site
impurity is
given by $\varepsilon/v_F$ and for the hopping impurity by $1-b$.
As discussed in the Appendix $a_2(V) (2-2K) \geq 0$.

For {\it repulsive} interaction ($K<1$) we expect $\chi$ to increase as a
power law. This power law holds for $M$ with $(W_b/v_F)^2 M^{2-2K} \ll 1$
but $M \gg 1$. For larger $M$ the expansion Eq.\ (\ref{backwardo})
breaks down and for smaller $M$ the corrections symbolized by the dots
become important.
Fig.\ \ref{fig4}
shows a double logarithmic plot of $\chi$ as a function of $M$
for weak hopping and site impurities with different
$b$, $\varepsilon$ and $V$.
The lines are power laws with the expected exponent $2-2K$.
For $V=0.445$ Eq.\ (\ref{eqn2a}) gives $2-2K=0.25$ and for $V=1$,
$2-2K=0.5$.
For large $M$ the DMRG results approach the predicted power law
behavior.  As we
have chosen small $1-b$ and $\varepsilon/|t|$, the power law holds for all
accessible system sizes with $M>40$.

\begin{figure}[tb]
\begin{center}
\leavevmode
\epsfxsize \columnwidth
\epsffile{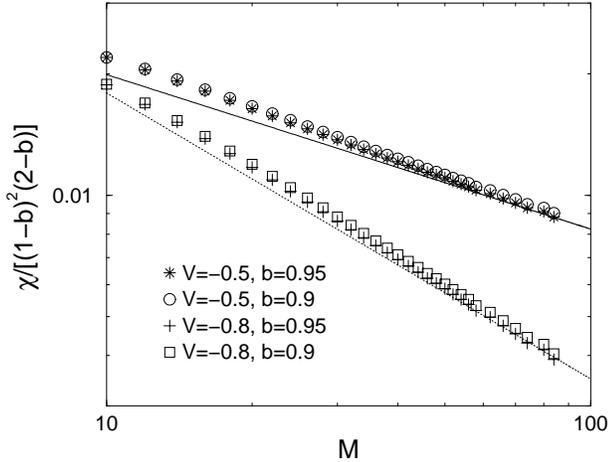}
\caption{Scaling behavior of the overlap for weak hopping
impurities and attractive
interaction.
The symbols are the DMRG results for parameters as indicated in
the legend.
The solid line presents a power law with
exponent $2-2K=-0.3834$  and the dotted line has the exponent $2-2K=-0.7100$.
In the legend $V$ is measured in units of $|t|$.}
\label{fig5}
\end{center}
\end{figure}

For {\it attractive} interaction ($K>1$) and weak hopping
impurities Eq.\ (\ref{plottingway1})
predicts that $\chi$ tends to zero as $M^{2-2K}$.
This power law behavior should
hold as long as $2-2K>-1$, \ie for $1<K<3/2$.
For larger $K$ the $1/M$ correction included
in the terms symbolized by the dots dominates the large $M$ behavior.

If we calculate $\chi/(W_b/v_F)^2$ the right hand side (rhs) of
Eq.\ (\ref{plottingway1}) is independent of $(W_b/v_F)^2$,
i.\ e.\ we expect the data for large $M$, fixed $V$ but
different $W_b$ to collapse onto one curve.
The lowest order expression of $W_b/v_F$ in terms of the
microscopic impurity parameter is given by $1-b$.
We calculated  $\chi/(1-b)^2$
for $V/|t|=-0.5$ and $V/|t|=-0.8$ each for $b=0.95$ and $b=0.9$.
From Eq.\ (\ref{eqn2a}) we obtain for
$V/|t|=-0.5$, $2-2K=-0.3834$ and for  $V/|t|=-0.8$,
$2-2K=-0.7100$.
Even though $1-b$ is very small the data points for the two
different $b$ do {\it not} collapse onto one curve and we find a
deviation between the power law behavior we expect from
Eq.\ (\ref{plottingway1}) and the behavior of the data at
large $M$ (see Fig.\ \ref{fig5}).
This shows that {\it higher order corrections} are of importance.
From a numerical point of view it is impossible to work with
smaller $1-b$, as the maximal difference between the overlap $O$ and
$1$ for $b=0.95$ is only $1.5 \cdot 10^{-4}$. For smaller $1-b$ this
difference would be even smaller and we would get very close to the
numerical accuracy of $O$.

We have to distinguish between two different sources of higher order
corrections.
The first of these arises in the mapping from microscopic to
field theoretical impurity parameters.
As discussed in the context of Eq.\ (\ref{weak}),
the higher order corrections to $\alpha^B(V=0)$
for the hopping impurity are of the order $(1-b)^3$.
If we include this correction and make the identification
\begin{eqnarray}
\label{identification}
\frac{W_b^2}{v_F^2} = (1-b)^2 \left[ 1+ (1-b) \right]
\end{eqnarray}
the different data sets are only {\it weakly} $b$ dependent.
Even if we include these corrections the results for different $b$
do not approach one another for large $M$. Instead the difference
between the $b=0.95$ and $b=0.9$ data sets slowly increases as $M$
gets larger. This can be explained by higher order corrections
in Eq.\ (\ref{backwardo}) (see below).
Fig.\ \ref{fig5} shows a double logarithmic plot
of $\chi/[(1-b)^2(2-b)]$ for the above parameters.
We also calculated $\chi/\delta_b^2$ with the exact
non--interacting backward scattering phase shift $\delta_b$ given
by Eq.\ (\ref{bialpha}), but this gives no further improvement compared
to the results shown in Fig.\ \ref{fig5}.
We therefore conclude that even for
relatively small $1-b$ it is necessary to take higher order
corrections in the mapping of the microscopic
impurity parameters onto the field theoretical parameters into account.

The slight deviation of the numerical data from the  expected asymptotic
power law behavior (shown by the lines in Fig.\ \ref{fig5})
suggests that higher order corrections are {\it also}
important in the field theoretical calculation leading
to Eq.\ (\ref{backwardo}).  Unfortunately it is difficult to
calculate these corrections analytically, even within the TL model.
As, among the cases considered here, these effects seem to be
important only for weak hopping impurities with attractive interaction,
we will not pursue this question further in the present article.

We also performed DMRG calculations for weak bond impurities. The results
are equivalent to the results we obtained for hopping impurities.

In summary, we conclude that our highly accurate DMRG data show
the expected weak impurity behavior in most of the cases discussed.
For site impurities and attractive
interaction we have numerically verified the Eq.\ (\ref{wf}) for
$\alpha_f^B(V)$. The backward scattering contribution
to $\alpha$ does vanish. For weak site, hopping, and bond impurities
with repulsive interaction the data are in agreement with the scaling
law Eq.\ (\ref{backwardo}). Only for weak hopping and bond impurities
and {\it attractive interaction} we found small deviations between the
scaling law  Eq.\ (\ref{backwardo}) and the numerical results. In this case
higher order corrections are of importance.

\subsection{Weak hopping in an open chain}
\label{weakhopsec}

To verify Eq.\ (\ref{backwardodual}) for our microscopic lattice model
we calculated the overlap $O^o$ between
the chains with OBC (hopping impurity with $b=0$)
and with a strong hopping impurity, \ie
small $t_b$, for repulsive
interaction.  There is no forward scattering contribution to the
overlap and $O^o$ is given by $O_b^o$ alone.
Again it is very important to have numerical data for $O^o$
with a large accuracy
as the difference between $O^o$ and $1$ is of the order of $10^{-4}$.

\begin{figure}[tb]
\begin{center}
\leavevmode
\epsfxsize \columnwidth
\epsffile{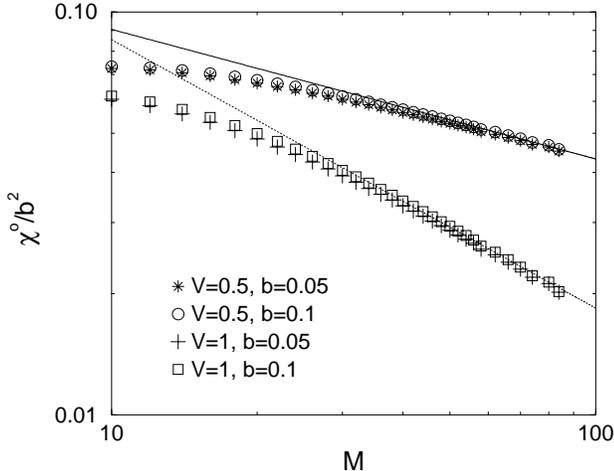}
\caption{Scaling behavior of the overlap $O^o$ between the open chain and
a chain with a weak hopping for repulsive interaction.
The symbols are the DMRG results for parameters as indicated in
the legend.
The solid line is a power law with
the expected exponent $2-2/K=-0.3217$ and the dotted line has exponent
$2-2/K=-2/3$. In the legend $V$ is measured in units of $|t|$.}
\label{fig6}
\end{center}
\end{figure}

The constant $\tilde{a}_1(V)$ in
Eq.\ (\ref{backwardodual}) is unknown and as above we
analyze $O^o$ by calculating centered differences. For these we expect
\begin{eqnarray}
\label{plottingway2}
\chi^o[i] & \equiv & -
\frac{O^o(M[i+1])-O^o(M[i-1])}{  \ln{ \left\{ M[i+1] \right\} }
-     \ln{ \left\{ M[i-1] \right\} } } \nonumber \\
& = &
\tilde{a}_2(V) \frac{t_b^2}{t^2}
(2-2/K) M^{2-2/K} + \ldots \,\, ,
\end{eqnarray}
to hold. As in the weak impurity case
we have $\tilde{a}_2(V) (2-2/K) \geq 0$.
Provided that $2-2/K > -1$ and $K<1$ (\ie $2/3<K<1$) $\chi^o[i]$
decays as a power law in $1/M$, with exponent $2-2/K$.  For smaller $K$
the $1/M$ correction to Eq.\ (\ref{plottingway2}) (among the terms
indicated by the dots), dies away more slowly than the $M^{2-2/K}$
term, and dominates the behavior at large $M$.  Furthermore
if we divide both sides of
Eq.\ (\ref{plottingway2})  by $(t_b/t)^2$ the rhs
is independent of $t_b$. For the weak hopping case
$(t_b/t)^2$ is given by $b^2$.
Therefore Eq.\ (\ref{plottingway2}) predicts that
the data for large $M$,
different $b$, but the same $V$ collapse onto the same
curve.
Fig.\ \ref{fig6} shows a plot
of $\chi^o/b^2$ as a function of $M$ for $V/|t|=0.5$ and $V/|t|=1$,
each for $b=0.05$ and $b=0.1$. The analytical exponents $2-2/K$
for these $V$ are $-0.3217$ and $-2/3$. The lines are power laws
with the expected exponents.

As predicted by perturbation theory,
the data for different $b$ but the same $V$ asymptotically
collapse onto one curve.
For large $M$ these curves show
power law behavior with the expected exponent $2-2/K$.
The numerical results are consistent with Eq.\ (\ref{backwardodual}),
i.\ e.\ the OE $\alpha^o$ tends to zero and the OBC ground state and
the ground state with a weak hopping are contrary to the non--interacting
case not orthogonal to each other.

\subsection{Strong impurities}
\label{strongsec}

The main question that we will address in this section is
whether the backward scattering contribution to the OE for repulsive
interaction tends to $1/16$ as $M \to \infty$, i.\ e.\ the open chain
interpretation is valid.
To avoid complications due to the forward scattering contribution
to $\alpha$ we concentrate on the hopping and bond impurities.
Results for strong site impurities will be discussed elsewhere
\cite{secondpaper}.

In Ref.\ \cite{qinI} Qin et al.\  also discuss the question of
whether $\alpha_b$ tends to $1/16$ for a
{\it bond impurity} with $b=0.1$, $V/|t|=1$ and up to $48$ sites
by plotting
$\alpha^O[i]$ (see Eq.\ (\ref{centerddef})).  As described in
Sec.\ \ref{secextra}
the extrapolation error in $\alpha^O[i]$ for a bond impurity
is much larger than that for a site impurity, \ie the finite size
corrections
are much more important.  Because of the anomalous corrections
expected for interacting electrons (see Eq.\ (\ref{backwardo}))
we cannot extrapolate by fitting a polynomial in $1/M$.
It is therefore questionable whether we can extract the $M \to \infty$
limit by plotting the bare data for $\alpha^O[i]$.

\begin{figure}[tb]
\begin{center}
\leavevmode
\epsfxsize \columnwidth
\epsffile{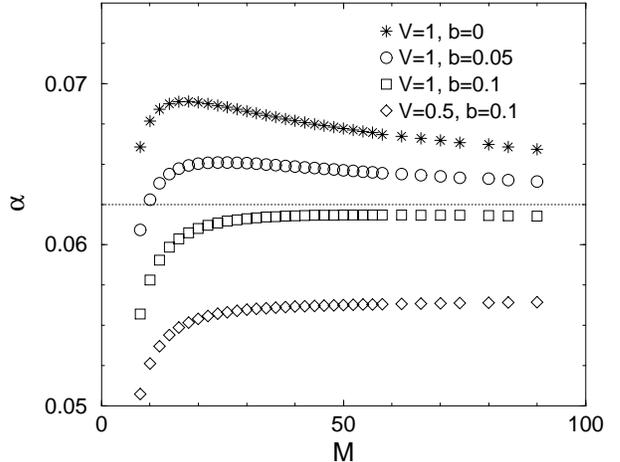}
\caption{Behavior of $\alpha^O[i]$ for a  repulsive interaction in the
limit
of strong bond impurity.
The symbols are the DMRG results for parameters as indicated in
the legend.
The dotted line is $\alpha=1/16$.
In the legend $V$ is measured in units of $|t|$.}
\label{fig7}
\end{center}
\end{figure}

Fig.\ \ref{fig7} shows $\alpha^O[i]$ for several $V$ and $b$.  The squares
indicate data calculated using the same parameters as used in Ref.\
\cite{qinI}, but for much
larger system sizes.  Provided that the open chain interpretation
of the RG and perturbative
arguments is correct, all the curves should converge
to $1/16$ in the limit $M \to \infty$.
Contrary to Qin et al.\ we find that it is {\it impossible} to confirm
{\it or}
refute this prediction by plotting $\alpha^O[i]$.  Within the range
of system sizes accessible, {\it none} of the curves saturates at
$1/16$ --- finite $M$ corrections dominate the scaling for all values
of $b$.  We note that the errors of the data presented in Fig.\ \ref{fig7}
are in all cases smaller than the symbol size.
It is known from BCFT that $\alpha^O[i]$ approaches $1/16$ for $b=0$
\cite{affleck94},
but even this simple case is not obviously confirmed by Fig.\ \ref{fig7}.
The behavior of the data for the {\it hopping impurity} is similar.

We have tested various different kinds of extrapolation
procedure (\eg second order polynomials and
polynomials combined with terms which include anomalous scaling) as fits
to
the behavior of $\alpha^O[i]$ in the limit $M[i] \to \infty$, but none of
these proved satisfactory for either the bond or the hopping
impurities.  For $\alpha^O[i]$ the corrections to the $M \to \infty$
result are
too large. In the following we therefore concentrate on $\Delta E$.
This means that we from now on {\it assume} that we can use the BCFT result
Eq.\ (\ref{bcftenergy}) to calculate the OE $\alpha$.
In addition we will continue the analysis focusing on the
more physical {\it hopping impurity}.

At the RG fixed point $b=b^*=0$ the finite $M$ corrections to $\Delta E$ are
given by integer powers of $1/M$ \cite{affleck94,qinII}
\begin{eqnarray}
\label{fixexpan}
\Delta E (V,b^*)
& = & c_0
+ c_1(V,b^*) M^{-1}
\nonumber \\*
&& + c_2(V,b^*) M^{-2}
+ \ldots ,
\end{eqnarray}
with constants $c_i$, and the quadratic
extrapolation of $ \alpha^{\Delta E}[i]$ for $M[i] \to \infty$
can be applied (see  Eq.\ (\ref{centerddefa})).
We extrapolated $ \alpha^{\Delta E}[i]$  for the hopping impurity
as in the Figs.\
\ref{fig1} and \ref{fig3} and obtained
$\alpha^{\Delta E}_{\mbox{\scriptsize extr}}(V/|t|=0.5)=0.06252$,
$\alpha^{\Delta E}_{\mbox{\scriptsize extr}}(V/|t|=1)=0.06261$, and
$\alpha^{\Delta E}_{\mbox{\scriptsize extr}}(V/|t|=1.5)=0.06285$.
The absolute extrapolation error for non--interacting electrons and a strong
hopping impurity (discussed in Sec.\ \ref{secextra}) is of the order
of $10^{-5}$ (see Table \ref{table1}).
The extrapolated $\alpha$ for $V/|t|=0.5$
is consistent with the value $\alpha=1/16$, within this error.
For  $V/|t|=1$ and $V/|t|=1.5$ the difference between the
extrapolated $\alpha$ and
$1/16$ is slightly larger. Taking into account the error due to
the approximate
nature of the DMRG and the fact that the extrapolation error
might be larger for interacting electrons, we can still
conclude that the results
for hopping impurities with $b=0$ are consistent with $\alpha=1/16$. The same
holds for bond impurities.

For $b \neq 0$ we expect anomalous corrections in the finite size
behavior of $ \Delta E(V,b)$ and therefore to $ \alpha^{\Delta E}[i]$.
Unfortunately we cannot extract the anomalous scaling behavior
directly from our data for $ \alpha^{\Delta E}[i]$,
as it is masked by the normal behavior Eq.\ (\ref{fixexpan}).
From the RG flow equations given in Ref.\
\cite{kfc}, we expect that close to the $b^*=0$ fixed point the effective
$b$ is given by $b M^{-\eta}$, with $-\eta = 1-1/K$,
\ie $b$ flows to zero as we increase the system size.
The idea is now to replace $b^*$ in
Eq.\ (\ref{fixexpan}) by $b M^{-\eta} $ and expand
the difference between $ \Delta E(V,b)$ and  $ \Delta E(V,b^*)$ in
$b M^{- \eta}$.
Before doing this it is important to notice that the constant
$c_0 $ in
Eq.\ (\ref{fixexpan}) is a functional of the Fourier transform of
the impurity potential  $ \tilde{W}_b(k) $ and not only a function
of the effective backward scattering strength. Therefore we cannot
introduce
the effective $b$ in this term and the leading term in the difference
is an unknown constant $\tilde{c}_0$.
The expansion gives
\begin{eqnarray}
\label{diffexpan}
&& \Delta E(V,b) - \Delta E (V,b^*) =  \tilde{c}_0
+ b \left. \frac{\partial c_1(V,b)}{\partial b} \right|_{b=b^*}
M^{-1-\eta}
\nonumber \\
&& + b \left. \frac{\partial c_2(V,b)}{\partial b} \right|_{b=b^*}
M^{-2-\eta}  + \ldots  \nonumber \\
&& + \frac{1}{2} b^2
\left. \frac{\partial^2 c_1(V,b)}{\partial b^2} \right|_{b=b^*}
M^{-1-2\eta}   + \ldots \,\, .
\end{eqnarray}
Taking the derivative (numerically the centered differences)
of this expression with respect to $1/M$ we arrive at
\begin{eqnarray}
\label{diffexpana}
&& \Delta_E (V,b)  \equiv
\frac{ \partial
\left\{ \Delta E(V,b) - \Delta E (V,b^*) \right\} }{\partial (1/M)}
\nonumber \\
&&  =   b \left. \frac{\partial c_1(V,b)}{\partial b} \right|_{b=b^*}
(1+ \eta) M^{1-1/K}
\nonumber \\
&& + b \left. \frac{\partial c_2(V,b)}{\partial b} \right|_{b=b^*}
(2+\eta) M^{-1/K} + \ldots  \nonumber \\
&& + \frac{1}{2} b^2
\left. \frac{\partial^2 c_1(V,b)}{\partial b^2} \right|_{b=b^*}
(1+2 \eta) M^{2-2/K} + \ldots \,\, .
\end{eqnarray}
Provided we can use the scaling law for the effective $b$ 
in the derivation of Eq.\ (\ref{diffexpan}) 
we expect that our data can be described by Eq.\ (\ref{diffexpana}).

\begin{figure}[tb]
\begin{center}
\leavevmode
\epsfxsize \columnwidth
\epsffile{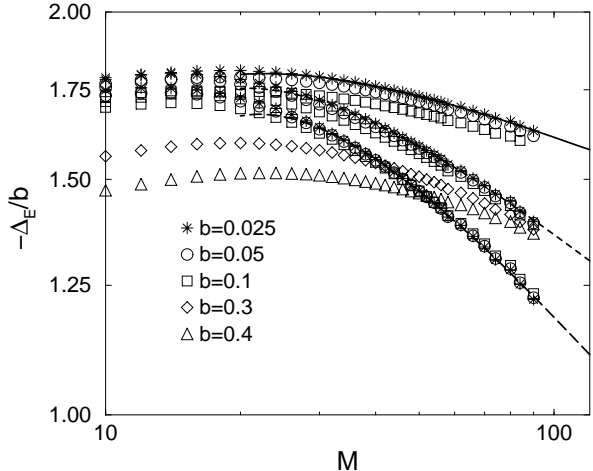}
\caption{
Large $M$ behavior of $-\Delta_E/(b|t|)$ for repulsive interaction
and a strong hopping impurity. For each interaction strength $V/|t|=0.5$,
$V/|t|=1$, and $V/|t|=1.5$ the figure shows data for $b=0.025$,
$b=0.05$, and $b=0.1$. The top set of curves is for $V/|t|=0.5$, the set
in
the middle for $V/|t|=1$, and the bottom set for $V/|t|=1.5$. The lines
are fits to the data (for details see the text). The diamonds and triangles
are data for $b=0.3$ and $b=0.4$ both for $V/|t|=1$. In the figure $\Delta_E$
is measured in units of $|t|$. 
}
\label{fig8}
\end{center}
\end{figure}

Fig.\ \ref{fig8} shows $-\Delta_E/(b|t|)$ as a function of $M$ for
a hopping impurity with
$V/|t| = 0.5$, $V/|t| = 1$ and  $V/|t| = 1.5$, for
$b=0.025$, $b=0.05$ and $b=0.1$ (for a discussion of the data
for $b=0.3$ and $b=0.4$ see below).  The data
for the different $V$ and large $M$ collapse onto a single curve.
As we expect from Eq.\ (\ref{diffexpana}),
the $b$ dependent corrections
[\eg the term in the fourth line of Eq.\ (\ref{diffexpana})] are
more important for weak interaction, \ie $K$ closer to $1$,
than for strong interaction.
It is obvious from Fig.\ \ref{fig8} that, for
the accessible system sizes, the numerical data cannot be described by
the leading term of Eq.\ (\ref{diffexpana}) alone.
Taking only the powers of $M$ into account the term proportional to
$b^2$ in Eq.\ (\ref{diffexpana}) seems to be a
more important finite size correction than that proportional
to $b M^{-1/K}$.  But as the data for small $b$ and larger
$V$ have already collapsed onto a single curve for $M>50$, we conclude
that the
term proportional to $b^2$ is already small and neglect this contribution
in what follows.
To verify that the data fall away as predicted
we fitted the points for $b=0.025$ and $M \geq 30$ with a function of
the form
\begin{eqnarray}
\label{fitform}
f(M) = d_1 M^{1-1/\tilde{K}} + d_2 M^{-1/\tilde{K}} ,
\end{eqnarray}
with fitting parameters $d_1$, $d_2$, and $\tilde{K}$.
The fits are shown as lines in Fig.\ \ref{fig8}.
The exact values for $K$ are $K=0.861$, for $V/|t| = 0.5$,
$K=3/4$, for $V/|t| = 1$, and
$K=0.649$, for $V/|t| = 1.5$. These numbers must be
compared with the exponents  $\tilde{K}=0.873$,  $\tilde{K}=0.774$,
and $\tilde{K}=0.706$ obtained by fitting the data.
For $V/|t| = 0.5$ and $V/|t| = 1$ the agreement between the fitted and the
expected $K$ is satisfactory. For large interaction the agreement is not
that good. In this case higher order corrections proportional to $b$
seem to be more important. If we fit the data in a smaller $M$ window,
\ie only for the few largest system sizes available,
the difference between the fitted and exact $K$ is decreasing as it should
be.

We conclude that for $M \to \infty$ the $M$ dependent part of 
$\Delta E(V,b)$ approaches the $M$ dependent part of the fixed
point energy difference $\Delta E(V,b^*=0)$ in the way expected from RG.  
In particular we find no sign of a term of order $M^0$ 
in $\Delta_E(V,b)$. 
If $\alpha^{\Delta E}(V,b)$ would be different
from $\alpha^{\Delta E}(V,b^*)$, $\Delta_E(V,b)$ would contain such a
constant given by $2 \pi v [\alpha^{\Delta E}(V,b)-  
\alpha^{\Delta E}(V,b^*)]$. By plotting $\Delta_E(V,b)$ as in Fig.\ \ref{fig8}
we are able to detect such a constant if it is up to $100$ times 
smaller than $2 \pi v_F [\alpha^{\Delta E}(V=0,b)-  
\alpha^{\Delta E}(V=0,b^*)]$.
As the numerical results for
$\alpha^{\Delta E}_{\mbox{\scriptsize extr}}(V,b^*)$ are
consistent with $\alpha=1/16$ it follows that $\alpha=1/16$ holds also
for {\it small} $b$ away from the fixed point value
{\it provided} that the use of BCFT predictions is valid.
We wish to emphasize that we can give
{\it quantitative} numerical evidence for the predicted RG flow
of the ground state energy difference for a fixed $b$, and not only a
{\it qualitative} picture of the flow by choosing different $b$ as was
obtained in Ref.\ \cite{qinII}.

We performed the same kind of analysis for the bond impurity. The results
are similar to those presented for the more physical hopping impurity and the
same conclusions can be drawn.

High accuracy DMRG then provides indirect numerical evidence that, for
repulsive interaction, the backward scattering
contribution to the OE for the overlap between the PBC ground state
and the state with a local impurity is independent of both interaction
and impurity strength, and has the universal value $1/16$.
This result is consistent with
those obtained by approximative analytical calculations
in the effective low energy field theory based on the open chain
interpretation and discussed in the introduction.

Even though we have studied chains with up to $100$ sites,
very small $b$ and obtain DMRG
data with a very high accuracy, we can only give indirect evidence for
this universal behavior
because finite size corrections are very large.  To show directly that
$\alpha^O[i]$ (and $ \alpha^{\Delta E}[i]$) tend to $1/16$ would require
much larger system sizes.

Fig.\ \ref{fig8} also contains data for  $b=0.3$, $b=0.4$ and $V/|t|=1$.
For the accessible system sizes the data points for these larger
values of $b$ are further away from the asymptotic curve onto which the
data for smaller $b$ collapse.
However the points {\it do} seem to converge with this curve for
large $M$.  One can therefore {\it speculate} that
the backward scattering contribution to the OE is also $1/16$
for {\it larger} values of $b$.
This is consistent with the open chain interpretation of the RG results.

\section{Summary}
\label{sum}

In this article we presented the results of an extremely
accurate DMRG study of
the orthogonality catastrophe in a 1D lattice model of correlated
electrons. We considered three different types of impurity and a range
of values of the interaction. By making a detailed comparison with exact
results for non--interacting electrons we were able to establish the accuracy
of our DMRG data, and arrive at a method of extrapolating our finite size
results to the thermodynamic limit.

For {\it weak} impurities we found that, in most of the cases considered,
the numerical data can be well described by the results of a 
analytical calculation 
in lowest order perturbation theory in the
impurity strength.  The perturbation theory is performed
within the effective low energy continuum field theory for 1D correlated
electrons.  Higher order corrections are important
in the case of weak hopping and bond impurities with attractive interaction.
We have confirmed numerically that a {\it weak} backward scattering component
to the OE scales to zero for attractive interaction and that the forward
scattering contribution is modified by the interaction.  The data
of the overlap also
show an interaction dependent anomalous scaling
with system size which was predicted by perturbation theory in the
(backward scattering) impurity strength.

In the case of a {\it strong} impurity we were able to confirm
{\it indirectly} the
RG prediction that the backward scattering contribution to the OE tends
to $1/16$ in the thermodynamic limit. This value is universal and
independent of both impurity strength and the size of the interaction.
Contrary to the conclusion drawn in previous studies by Qin et al.\
we found that {\it direct} evidence for this behavior cannot be extracted
from data for the overlap at the accessible system sizes, because
the finite size corrections are very large.
However we obtained {\it quantitative} numerical evidence for the scaling
of the ground state energy difference predicted by RG.

The open chain interpretation of perturbative RG results leads to the
prediction that $\alpha_b(V>0)$ scales to $1/16$ in the presence of an
impurity with an {\it arbitrarily weak} backward scattering
component (e.\ g.\ a hopping or bond impurity with $b \geq 0.9$).
By close analogy $\alpha_b(V<0)$ should then tend to $0$ even
for very {\it strong} bare backward scattering (e.\ g.\ a hopping or
bound impurity with $b \leq 0.1$).
We found it {\it impossible} to confirm these predictions {\it quantitatively}
on the basis of DMRG data for systems with up to $100$ sites.
The finite size corrections are in all cases too large, and the scaling
too slow. {\it Much} larger systems are needed to make a reliable numerical
determination of these issues possible.
The results are however in {\it qualitative} agreement with this
extension of the RG results.

\section*{Acknowledgments}
We would like to thank the ISI Foundation
for making possible the workshop on ``The Role of
Dimensionality in Correlated Electronic Systems'' at Villa Gualino
(Torino, Italy) in May 1996 where this work was begun.
We are also grateful to the EU (HC\&M Network ERBCHRX--CT920020) for
financial support during our stay in Torino.  Most of the numerical
calculations were done on the IBM SP2 cluster at Indiana University.
We would like to thank the University Computing Services for support.
Additional calculations were done on the IBM SP2 at the
Leibniz--Rechenzentrum in Munich. We are grateful to C.\ Schuster
for useful comments on the manuscript. 

V.\ M.\ would like to thank K.\ Sch\"onhammer, S.\ Girvin, W.\ Zwerger,
S.\ Eggert, W.\ Atkinson, and U.\ Z\"ulicke for discussions and
the Deutsche Forschungsgemeinschaft and the NSF Grant No.\ DMR--9416906
for financial support.
This work was partially supported through the TMR program of the European
Union (P.\ S.).
N.\ S.\ would like to acknowledge useful conversations with
F.\ E{\ss}ler, B.\ Muzikanski, and N.\ d'Ambrumenil, and support
under an EPSRC grant.

\section*{Appendix}
\label{appendix}

In this Appendix we will briefly describe how to obtain the perturbative
expressions Eqs.\ (\ref{backwardo}) and  (\ref{backwardodual}) for the
overlap \cite{ks2}.
By ${\mathcal H}_0$ we denote a general Hamiltonian
and by ${\mathcal W}$ the perturbation. The overlap between the ground
state
$\left| E_0 \right>$ of ${\mathcal H}_0$ and $\left| E_0^I \right>$
of ${\mathcal H}_0 + {\mathcal W}$ is given by
\begin{eqnarray}
\label{ap1}
O^2 \equiv \left| \left< \left. E_0 \right| E_0^I \right> \right|^2 = 1-
\sum_{n \neq 0} \left| \left< \left. E_0 \right| E_n^I \right> \right|^2 ,
\end{eqnarray}
where $\left|  E_n^I \right>$ are the excited eigenstates of
${\mathcal H}_0 + {\mathcal W}$. Inserting the expression for
 $\left|  E_n^I \right>$ in lowest order perturbation theory in
${\mathcal W}$ gives
\begin{eqnarray}
\label{ap2}
O^2  \approx 1-
\sum_{n \neq 0} \frac{ \left| \left<  E_0 \left| {\mathcal W}
\right|  E_n \right> \right|^2}{\left( E_n -E_0 \right)^2} .
\end{eqnarray}
By introducing a $\delta$ function
and taking into account that the smallest energy difference $ E_n -E_0$
is proportional to $1/L$ we end up with
\begin{eqnarray}
\label{ap3}
O^2 \approx 1- \int_{2 \pi /L}^{\infty} \frac{d \nu}{\nu^2}
 \left<  E_0 \left| {\mathcal W}  \delta \left( \nu - \left[
{\mathcal H}_0 - E_0 \right] \right) {\mathcal W}  \right|  E_0 \right>
\end{eqnarray}

We will now specify ${\mathcal H}_0$ and  ${\mathcal W}$ to be given by
the Eqs.\ (\ref{fieldh}) and (\ref{hback}). In this case
the  correlation function
occurring in Eq.\  (\ref{ap3}) is related to the $2 k_F$ component
of the density--density correlation function in the TL model.
For the TL Hamiltonian it is known how to calculate
correlation functions like the one occurring in Eq.\ (\ref{ap3})
\cite{johannes}. Evaluating the expression gives
\begin{eqnarray}
\label{ap4}
O_b^2 & \approx & 1- c_1(V) \frac{W_b^2}{v_F^2} \nonumber \\
&& - c_2(V) \frac{2}{4 \pi^2}
\frac{W_b^2}{v_F^2}
\frac{L^{2-2K} - 1}{2-2K}
\end{eqnarray}
with positive constants $c_i$. An explicit expression
for $c_2(V)$ can be given, but
in the following we only need to know that $c_2(V=0)=1$ \cite{diplom}.
Eq.\ (\ref{ap4}) is equivalent
to Eq.\ (\ref{backwardo}). In the limit $K \to 1$, \ie  $V \to 0$,
$\left(L^{2-2K} - 1\right)/(2-2K)$ tends to $\ln{(L)}$
and we recover Eq.\ (\ref{backwardonointer}).

Taking the dual Hamiltonian describing an open chain
and a representation of the weak hopping in terms
of the dual field theory \cite{kfc} instead of
Eqs.\ (\ref{fieldh}) and  (\ref{hback})
we can perform  a similar kind of calculation and end up with
Eq.\ (\ref{backwardodual}).

In the non--interacting case we can furthermore write down higher order
corrections. They sum up to the expected
$O_b \sim \exp{ \left\{ - \alpha_b \ln{(L)} \right\} }$ behavior and
finite size corrections that are integer powers of $1/L$.

\clearpage

\widetext

\begin{table}
\begin{tabular}{cccccc}
   &  exact  & error from $O$  &  error from $\Delta E$  &
error from $O$  &  error from $\Delta E$  \\
 & & $M \in [200,1200] $ & $ M \in [200,1200] $
& $M \in [50,200] $ & $ M \in [50,200] $ \\
\\
site  & $4.89322299 \cdot 10^{-2}$ &  $ 1 \cdot 10^{-7} $ &
$ 1 \cdot 10^{-9} $ &   $ 2 \cdot 10^{-5} $ &   $ 5 \cdot 10^{-7} $ \\
 \\
hopping &  $4.76437497 \cdot 10^{-2}$ &  $ 5 \cdot 10^{-5} $ &
$ 3 \cdot 10^{-7} $ &   $ 2 \cdot 10^{-4} $ &   $ 1 \cdot 10^{-5} $
\end{tabular}
\caption{The first column gives the exact value of the OE $\alpha$
in the non--interacting tight binding Hamiltonian for
a site impurity with $\varepsilon/|t|=3$ and a hopping impurity with
$b=0.1$.
The other columns are the errors obtained by extrapolating
the data of Fig.\ 1.
The second and third  columns are the errors for an extrapolation with
$M \in [200,1200]$ and the fourth and fifth the ones for $M \in [50,200]$.}
\label{table1}
\end{table}

\begin{table}
\begin{tabular}{cccccc}
&$\alpha^{\mbox{\scriptsize exact}}(0)$
&$\alpha^B(0)$
&$\alpha_f^B(V)$
&$\alpha^{O}_{\mbox{\scriptsize extr}}(V)$
&$\alpha^{\Delta E}_{\mbox{\scriptsize extr}}(V)$     \\ \\
$V=-0.75$, $\varepsilon=0.05$
&  $ 3.1650 \cdot 10^{-5}$
&  $ 3.1663 \cdot 10^{-5}$
&  $ 3.7789 \cdot 10^{-5}$
&  $ 3.8152 \cdot 10^{-5}$
&  $ 3.9189 \cdot 10^{-5}$ \\ \\
$V=-1$, $\varepsilon=0.25$
&  $ 7.8342 \cdot 10^{-4}$
&  $ 7.9157 \cdot 10^{-4}$
&  $ 1.4072 \cdot 10^{-3}$
&  $ 1.3862 \cdot 10^{-3}$
&  $ 1.3991 \cdot 10^{-3}$ \\ \\
$V=-1.5$, $\varepsilon=0.1$
&  $ 1.2644 \cdot 10^{-4}$
&  $ 1.2665 \cdot 10^{-4}$
&  $ 7.4598 \cdot 10^{-4}$
&  $ 7.4128 \cdot 10^{-4}$
&  $ 7.4294 \cdot 10^{-4}$
\end{tabular}
\caption{The OE for weak site impurities. The first column gives the
exact non--interacting $\alpha$, the second column the non--interacting
Born approximation, the third column the forward scattering contribution
of the interacting Born approximation,
and the fourth and fifth columns
the extrapolated values from Fig.\ 3. The extrapolation is done by a
quadratic
fit with  $1/M \in [0.01,0.03]$.
In the Table $V$ and $\varepsilon$ are measured in units of $|t|$.}
\label{table2}
\end{table}

\end{document}